\documentstyle[aps,twocolumn,epsf]{revtex}
\begin{document}
\draft
\wideabs{
\title{Dynamics of two interacting Bose-Einstein condensates}
\author{A. Sinatra${}^1$, P.O. Fedichev${}^{2,3}$, 
Y. Castin${}^1$, J. Dalibard${}^1$, and G.V. Shlyapnikov${}^{1,2,3}$}
\address{${}^1$ Laboratoire Kastler Brossel$^*$,
24, Rue Lhomond, F-75231 Paris Cedex 05, France \\
${}^2$ FOM Institute for Atomic and Molecular Physics, Kruislaan 407, 1098 SJ
Amsterdam, The Netherlands \\
${}^3$ Russian Research Center, Kurchatov Institute, 
Kurchatov Square, 123182 Moscow, Russia}
\date{\today}
\maketitle

\begin{abstract}
We analyze the dynamics of two trapped interacting Bose-Einstein
condensates in the absence of thermal cloud 
and identify two regimes for the evolution: a regime of slow 
periodic oscillations and a regime of strong non-linear mixing leading to the 
damping of the relative motion of the condensates. 
We compare our predictions with an experiment recently performed at JILA.
\end{abstract}
}
\narrowtext
The experimental evidence for Bose-Einstein condensation
in trapped atomic gases \cite{BEC} has attracted a lot of attention, 
as the presence of a macroscopically occupied quantum state
makes the behavior of these gases drastically different
from that of ordinary gas samples. Trapped Bose-Einstein condensates are
well isolated from the environment and, at the same time, can be 
excited by deforming the trap or
changing the interparticle interaction. The question of how  
the gas sample, being initially a pure condensate,
subsequently reaches a new equilibrium state is directly 
related to the fundamental problem of the appearance of irreversibility 
in a quantum system with a large number of particles. 
Thus far the time dependent dynamics of trapped condensates has mainly
been analyzed for a single condensate 
\cite{Rup,Kagan,Castin,Smerzi,Holland} on the basis of 
the Gross-Pitaevskii equation for the condensate wavefunction. 
Remarkably, already in this mean field approach the stochastization in the
condensate evolution has been found \cite{Kagan}, and the  damping
of the condensate oscillations has been observed numerically \cite{Smerzi}. 
However, the question of the formation of a thermal component,
addressed in \cite{Kagan}, has not been investigated.

In this paper we study  
the evolution of a richer system, a mixture of two interacting condensates
($a$ and $b$), in the situation where initially the thermal cloud is absent.
The properties of a static two-component trapped condensate,
including the issue of spatial separation of the $a$ and $b$ components due
to interparticle interaction \cite{Fetter,Meystre}, were 
investigated in \cite{2comp}. The response of the system to small
modulations of the trap frequency has also been studied numerically \cite{Big}.
In our case the $a$ and $b$ condensates have initially the same density 
profile and are set into motion
mostly by an abrupt displacement of the trap centers. 
The main goal of our work is to study the dynamics of spatial separation of
the two condensates and analyze how the system can acquire statistical
properties and reach a new equilibrium state. 
 From a general point of view, 
we are facing the problem  
raised by Fermi, Pasta and Ulam \cite{Fermi}. 
They considered classical vibrations of a chain of 
coupled non-linear oscillators, 
to analyze the emergence of statistical properties in a 
system with a large number of degrees of freedom. 
As has been revealed later, the appearance of statistical properties requires a 
sufficiently strong non-linearity leading to stochastization of motion 
\cite{Chir}, whereas for small non-linearity the motion remains quasiperiodic 
(see e.g. \cite{Sagd}).

We consider a situation in which the two condensates $a$ and $b$ see
harmonic trapping potentials of exactly the same shape, and the 
interparticle interactions characterized by the scattering lengths $a_{aa}$,
$a_{ab}$ and $a_{bb}$ are close to each other.
The control parameter, determining the possibilities of 
non-linear mixing and 
stochastization, is the relative displacement $z_0$ of the 
trap centers. 
We identify two regimes for the evolution. In the first one the relative
motion of the condensates exhibits oscillations at a frequency much
lower than the trap frequency $\omega$.
In the other regime there is a strong non-linear mixing leading to the damping
of the relative motion, and the system has a tendency to approach a new 
equilibrium state.   
We compare our predictions with the results of the JILA 
experiments \cite{Cornell1,Cornell2} on a two-component condensate of
${}^{87}$Rb atoms in the $F=1,m=-1$ and $F=2,m=1$ states. 
In these experiments the double condensate was prepared from a single 
condensate in the state $F=1,m=-1$ ($a$) by driving a two-photon 
transition which coherently transfers half of the atoms to the state 
$F=2,m=1$ ($b$).

We mostly perform our analysis in the mean field approach relying on the 
Gross-Pitaevskii equations for the wavefunctions $\phi_a$ and $\phi_b$ of the 
$a$ and $b$ condensates.
This approach corresponds to the classical limit of the evolution of a quantum 
field, the subsequent corrections being proportional to a small parameter 
$(na_{\varepsilon\varepsilon^\prime}^3)^{1/2}$ ($n$ is the gas density) and, hence, manifesting themselves
only on a rather large time scale. 
The two coupled Gross-Pitaevskii equations for $\phi_a$ and $\phi_b$ normalized to unity
read
\begin{equation}
i\hbar\partial_t\phi_\varepsilon\! =\!
\left[-{\hbar^2 \Delta\over 2m} + U_\varepsilon - \mu +\!\! \sum_{\varepsilon'=a,b}
\!g_{\varepsilon\varepsilon'} N_{\varepsilon'} |\phi_{\varepsilon'}|^2 \right]\!
\phi_\varepsilon.
\label{eq:GPE}
\end{equation}
Here $g_{\varepsilon\varepsilon'}=4\pi\hbar^2a_{\varepsilon\varepsilon'}/m$ are the 
coupling constants for elastic interaction between atoms in the states
$\varepsilon$ and $\varepsilon'$, $m$ is the atom mass, and $N_{\varepsilon}$,
$U_\varepsilon$ are the number of atoms and trapping potential for the
$\varepsilon$ condensate.
As in the JILA experiment, 
we choose the initial condition $\phi_{a,b}(0)=\phi_0$, where the (real)
wavefunction $\phi_0$ corresponds to the ground state of Eq.(\ref{eq:GPE})
with all atoms in the $a$ state and no trap displacement. 
The chemical potential of this ground state is denoted as $\mu$. 

We consider the $a$ and $b$ condensates in the Thomas-Fermi regime
($\hbar\omega\!\ll\!\mu$) and assume the number of condensate atoms
$N_a\!=\!N_b\!=\!N/2$ \cite{note}. 
The first set of our calculations is performed for the evolution
of the condensates in a spherically symmetric trapping potential 
$U_0(r)\!=\!m\omega^2r^2/2$ which at $t\!=\!0$ is displaced along the $z$ axis
by a distance $z_0/2$ for the $a$ atoms, and by $-z_0/2$ for the $b$ atoms.
We present the results for the time dependence of the mean 
separation between the condensates, 
\begin{equation}       \label{u}
u(t)=\int d^3r\;z\; (|\phi_a({\bf r},t)|^2-|\phi_b({\bf r},t)|^2).
\end{equation}
For the curves in Fig.1 the coupling constants are $g_{aa}=g_{ab}=g_{bb}$,
and for $z_0=0$ our initial state is an 
equilibrium state at $t\geq 0$.
In this state the Thomas-Fermi radius of the condensate $R_0=(2\mu/m\omega^2)^{1/2}$ 
serves as unit of length, and the shape of $\phi_0$ is determined by 
$\mu/\hbar\omega$.
Hence, for $z_0\neq 0$ the dependence of the quantity $u/R_0$ on 
$\omega t$  is governed by the parameters $\mu/\hbar\omega$ and $z_0/R_0$.

\begin{figure}
\epsfxsize=\hsize
\epsfbox{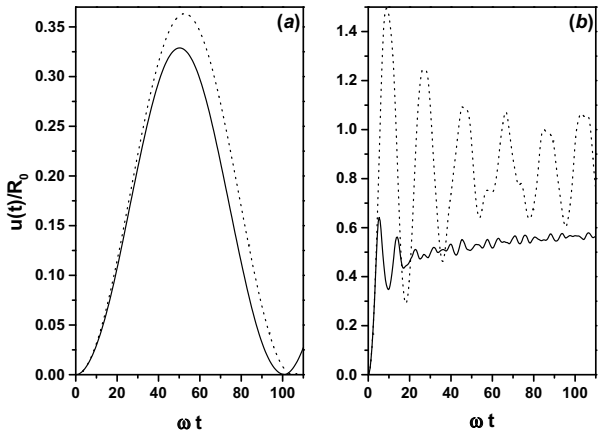}
\caption{ \protect
Mean separation between the condensates versus time in isotropic traps 
for $g_{aa}=g_{ab}=g_{bb}$ and $\mu/\hbar\omega=29.2$. 
Relative displacement: $z_0=6.66\times 10^{-4}R_0$ (a), 
and $z_0=7.16\times 10^{-2}R_0$ (b).
Solid curves: numerical integration of Eq.(\ref{eq:GPE}). 
Dotted curves: analytical prediction for (a) (see text), and the linear model
relying on Eq.(\ref{lin}) for (b). 
}
\label{1}
\end{figure}

Our results reveal two key features of the evolution dynamics. The first one,
for a tiny displacement $z_0$, is a periodic 
motion with slow frequencies which turn out to be sensitive to small 
variations in the values of the coupling constants. The other feature,
for much larger $z_0$, but still $z_0 \ll R_0$, 
is a strong damping in the relative motion of the two 
condensates, as observed at JILA \cite{Cornell1}.

In order to understand the physics behind the evolution pattern, we first 
perform a linear analysis of Eq.(\ref{eq:GPE}).
For the case where $g_{aa}\!=\!g_{ab}\!=\!g_{bb}\!=\!g$, and the displacement 
$z_0$ is sufficiently small, we linearize Eq.(\ref{eq:GPE}) with respect to 
small quantities $\delta\phi_{a,b}\!=\!(\phi_{a,b}\!-\!\phi_0)$ and $z_0$.
Then, for the quantity $\delta\phi_-\!=\!\delta\phi_a\!-\!\delta\phi_b$,  
describing the relative motion of the condensates, we obtain the equation
\begin{equation}         \label{eq:diff}
i\hbar\partial_t\delta\phi_-=\left[-\frac{\hbar^2\Delta}{2m}+U_0-\mu+Ng
\phi_0^2\right]\delta\phi_-+S_-,
\end{equation}
with the source term $S_-\!=\!m \omega^2 z_0 z \phi_0$. 
For the quantity $\delta\phi_+\!=\!\delta\phi_a+\delta\phi_b$ we find an 
equation decoupled from $\delta \phi_-$ and 
without source terms. Hence, the initial condition $\delta\phi_+({\bf r},0)=0$ 
allows us to put $\delta\phi_+({\bf r},t)=0$ for $t\geq 0$.

For $S_-\!=\!0$ Eq.(\ref{eq:diff}) is 
the equation for the wavefunction of a particle moving
in the potential $V\!=\!U_0\!-\!\mu\!+\!Ng\phi_0^2$.
Stationary solutions of
this equation provide us with the eigenmodes of oscillations of the 
condensates with respect to each other.     
In the Thomas-Fermi limit the potential $V$, originating from the kinetic
energy of the condensate, is a smooth function of $r$  inside the condensate
spatial region $r<R_0$: 
$V=\hbar^2(\Delta\phi_0)/2m\phi_0\ll\hbar\omega$.
For $r>R_0$ this potential is close to $U_0-\mu$ and is much steeper. 
Replacing $V$ by an infinite square well of radius $R_0$ we obtain the 
energy spectrum of eigenmodes with large quantum numbers $n$: 
$E_{n,l}=(\pi\hbar\omega)^2(2n+l)^2/16\mu$, where $l$ is the orbital 
angular momentum. 
This explains the appearance of oscillations at a frequency much 
smaller than $\omega$ in our numerical calculations (see Fig.1a), since the
energy scale in the spectrum is $(\hbar\omega)^2/\mu\ll
\hbar\omega$. For the latter reason we call these eigenmodes soft modes.
Note that the soft modes for the relative motion of the two condensates
also exist in the spatially homogeneous case where they have a free-particle 
spectrum \cite{Fetter}.

As in our linear approach we have $\delta\phi_+({\bf r},t)=0$, Eq.(\ref{u}) 
for the mean separation between the condensates reduces to
$u(t)=2\int d^3r\, z\, \phi_0\, {\rm Re}\{\delta\phi_-\}$, and the contribution
to $u(t)$ comes from the components of $\delta\phi_-$ with $l=1,m_l=0$.   
Solving Eq.(\ref{eq:diff}) 
with the initial condition $\delta\phi_-({\bf r},0)=0$, we obtain $u(t)$
as a superposition
of components, each of them oscillating at an
eigenfrequency of a soft mode:
\begin{equation}
u(t)\!=\!z_0\!\sum_{n\geq 1}\!
\frac{2 m \omega^2}{E_{n1}}\left|\int\!\!d^3r\varphi_{n1}z
\phi_0\right|^2\left[
1\!-\!\cos\left(\!\frac{E_{n1}t}{\hbar}\right)\right]\!,
\label{eq:uexp}
\end{equation} 
where $\varphi_{n1}$ is the wavefunction of the soft mode with $l=1,
m_l=0$ and main quantum number $n$. 
Damping of oscillations of $u(t)$ could, in principle, originate from the 
interference between the components with different $n$ in Eq.(\ref{eq:uexp}). 
However, the source $S_-$ basically populates only 
the lowest soft mode, irrespective of the value of $z_0$: 
the amplitude of oscillations at the lowest eigenfrequency in 
Eq.(\ref{eq:uexp}) (the term with $n=1$) 
greatly exceeds the sum of the amplitudes of other terms. 
Hence, these oscillations remain undamped.
For the same reason their frequency and amplitude can be found with
$\varphi_{n1}$ replaced by the function $z\phi_0$ normalized to unity.
Using the Thomas-Fermi approximation for the condensate wavefunction 
\cite{TF}:  
$\phi_0^2(r)\!=\!15(1\!-\!r^2/R_0^2)/8\pi R_0^3$ for $r\!<\!R_0$, and
$\phi_0\!=\!0$ for $r\!>\!R_0$, we obtain $E_{11}\equiv\hbar\Omega=
(7/4)(\hbar\omega)^2/\mu$ which is very close to 
$E_{11}\!=\!1.62(\hbar\omega)^2/\mu$ calculated numerically.
Then, retaining only the leading term ($n=1$) in
Eq.(\ref{eq:uexp}), we find  
$u(t)\!\approx\!z_0(4\mu/7\hbar\omega)^2[ 1\!-\!\cos (\Omega t) ]$ shown in
dotted line in Fig.1a.
As one can see, the condition of the linear 
regime $u \ll R_0$ requires a very small displacement
\begin{equation}      \label{disp}
z_0 \ll (\hbar\omega/\mu)^2 R_0,
\end{equation}
and already a moderate $z_0$ as in Fig.1b 
is sufficient to drive the system out of the linear regime. 

We have performed a similar linear analysis for the case
where $g_{aa}\neq g_{ab} \neq g_{bb}$, but the relative
difference between the coupling constants is small.
Also in this case the source $S_-$ mostly generates 
oscillations of the condensates relative to each
other at a single frequency $\Omega'\ll
\omega$. For a relative difference between the coupling constants
much smaller than $(\hbar\omega/\mu)^2$, 
the frequency $\Omega'$ coincides
with the soft-mode frequency $\Omega$ found 
above. Otherwise the sign of 
$g_-=g_{aa}+g_{bb}-2g_{ab}$ becomes important. 
In particular, for positive $g_-\gg |g_{aa}-g_{bb}|$ already
a moderate difference between the coupling constants
strongly increases the
frequency $\Omega'$ compared to $\Omega$.
In this case we obtain 
undamped oscillations at $\Omega'\approx (g_-/g_{aa})^{1/2}\omega$.
For $g_-<0$, already in the $z_0=0$ case,
a breathing mode in which the two condensates oscillate
out of phase becomes unstable, and the system evolves far from the 
initial state.
Note that for a small difference between the coupling constants 
the condition $g_{-}<0$ is equivalent to the criterion of spatial
separation of the condensates
in the homogeneous case, $g_{aa}g_{bb}<
g_{ab}^2$ \cite{Fetter,Meystre}. 

We now turn to the large $z_0$ regime (Fig.1b) where
we find a strong damping of the oscillations of the mean separation
between the condensates, $u(t)$. In order to prove the 
key role of non-linearity 
in this regime, we first  
attempt a linear model assuming that the densities
$|\phi_{\varepsilon^\prime}|^2$ inside the square brackets of Eq.(\ref{eq:GPE}) are
not evolving:
\begin{equation}         \label{lin}
\sum_{\varepsilon^\prime}N_{\varepsilon^\prime}g_{\varepsilon\varepsilon^\prime}
|\phi_{\varepsilon^\prime}|^2\rightarrow
Ng|\phi_0|^2 \,. \nonumber
\end{equation} 
% In this model the two condensates evolve independently. 
In contrast to the analysis which led to Eq.(\ref{eq:uexp}), 
the displacement $z_0$ is now explicitly included in the Hamiltonian through the terms
$\pm m \omega^2 z z_0/2$ in $U_{a,b}$, and the number of populated
oscillation modes depends on $z_0$.   
However, for the parameters in Fig.1b we find that only a 
few modes are populated, and the interference between them can not 
account for the damping found numerically
(dotted versus solid curve in Fig. 1b). 

We argue that the damping in our calculations mostly originates from  
non-linearity of the system, which increases the number and amplitude 
of populated oscillation modes and provides an interaction between them. 
As a result, the evolution of the condensate wavefunctions $\phi_a$ and $\phi_b$ 
becomes chaotic.
This can be seen from Fig.2 where we compare   
the spectral density
$R_n(\nu)\!=\!|T^{-1}\int_0^T\!dt\,n({\bf 0},t)\exp{(i\nu t)}|^2$
of the density at the origin $n({\bf 0},t)$ with an identically defined 
spectral density $R_u(\nu)$ of $u(t)$ for the parameters in Fig.1b and 
$T=110/\omega$.
The function $R_n(\nu)$ has a smooth envelope at large $\nu$, 
with peaks corresponding to the islands of regular motion.
On the contrary, $R_u(\nu)$ exhibits pronounced peaks 
at $\nu$ of order $\omega$, without any smooth background.
This picture provides a clear signature of stochastization in the system 
\cite{Sagd} and prompts us to represent 
each of the condensate wavefunctions in Eq.(\ref{eq:GPE}) as a superposition 
of two constituents:
(i) a slowly oscillating regular part conserving the phase coherence properties;
(ii) a composition of high-energy excitations characterized by
stochastic motion.
Only the slow constituent contributes to such macroscopic quantities
as $u(t)$, since the contribution of the fast stochastic part is
averaged out. 

\begin{figure}
\epsfxsize=\hsize
\epsfbox{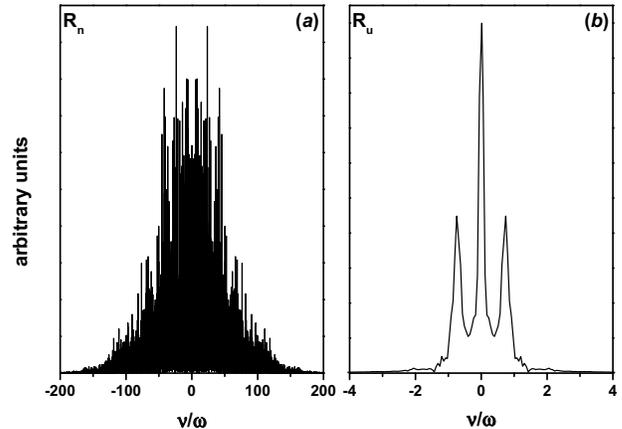}
\caption{ \protect
Spectral densities $R_n(\nu)$ (a) and $R_u(\nu)$ (b) 
for the parameters in Fig.1b and $T=110/\omega$ (see text).
}
\label{2}
\end{figure}

Our analysis is consistent with the general statement that for a large population 
of various oscillation modes the non-linear interaction between them leads to
stochastization in the motion of excitations with sufficiently high energy 
\cite{Sagd}. 
This allows us to employ the mechanism of stochastic heating 
\cite{Sagd} for explaining
the damping of oscillations of $u(t)$:
The mean field interaction between the fast stochastic 
and the slowly oscillating 
parts leads to energy transfer from the slow to the fast part.

The evolution of the occupation numbers of the modes of the fast stochastic part
is governed by kinetic equations
\cite{Sagd} and eventually slows down.
The rate of energy and particle exchange between the two 
constituents then
reduces. 
After a sufficiently long time only small linear oscillations of the condensates
survive, mostly  at the lowest eigenfrequency
and the gas sample as a whole could be thought as being close to
a steady state.
However the damping of the remaining oscillations and the ultimate evolution 
of the fast stochastic part towards the thermal equilibrium require 
an analysis beyond the mean field approach.
For the parameters in Fig.1b, using the semiclassical Bogolyubov approach
\cite{Stringari} and relying on the conservation of energy and number of
particles, we find an equilibrium temperature $T_{{\rm eq}}\approx 0.6\mu$ 
and a condensed fraction $\gamma_{a,b}\approx 0.9$, for $N=5\times 10^5$.

The last set of our calculations relates to the recent JILA experiment 
\cite{Cornell1}
where the evolution of a two-component $^{87}{\rm Rb}$ condensate has been 
investigated. In the conditions of this experiment we solved numerically
Eq.(\ref{eq:GPE}) by taking $a_{ab}=55$\AA$\,$ and the ratio 
$g_{aa}:g_{ab}:g_{bb}=1.03:1:0.97$.
We also explicitly included in these equations the 
$22$ ms expansion of the clouds after switching off the trapping potential.
The results of our calculations are presented in Fig.3.
As in Fig.1b, we find a strong damping of the oscillations of the mean
separation between the condensates, $u(t)$.
Our numerical results are in fair agreement with the experimental data, although
the damping in the experiment is somewhat larger.
We extended the calculations to twice the
maximum experimental time and found small oscillations which remain undamped
on this time scale.

\begin{figure}
\epsfxsize=\hsize
\epsfbox{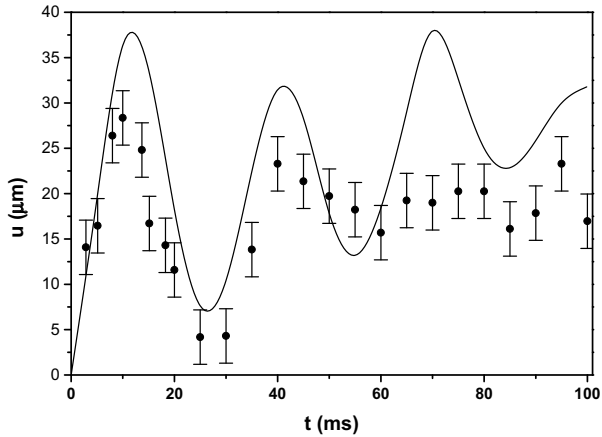}
\caption{ \protect
Mean separation between the condensates in the JILA 
experiment versus
evolution time in the traps, after a $22$ ms free expansion.
Dots with error bars: JILA experiment. Solid curve: our numerical calculation.
}
\label{3}
\end{figure}

Our data for the JILA experiment \cite{Cornell1} can be analyzed 
along the same lines as the results in Fig.1b, with a damping originating from 
stochastization in the evolution of the condensate wavefunctions.
The equilibrium temperature is close to $\mu$,
corresponding to condensed fractions 
$\gamma_a\approx\gamma_b\approx 0.9$. The large value of the condensed fraction explains why
phase coherence between the $a$ and $b$ components could be observed even after
the damping of the motion of $u(t)$ \cite{Cornell2}. 
The damping time of the small remaining oscillations,
estimated along the lines of \cite{FSW}, will be of order
  $1$ second.

We believe that the stochastic regime 
identified from our calculations is promising for investigating  
the loss of phase coherence and the formation of a new thermal component 
in initially purely Bose-condensed gas samples. 
An interesting possibility concerns the observation of a continuous change in the
phase coherence between the $a$ and $b$ components with increasing the trap 
displacement and, hence, decreasing the final Bose-condensed fraction.

This work was partially funded by EC (TMR network ERB FMRX-CT96-0002).
by the Stichting voor Fundamenteel Onderzoek der Materie (FOM), by the 
Russian Foundation for Basic Studies, by INTAS, and by NSF 
under Grant No. PHY94-07194. 

{$^*$ L. K. B.
is an unit\'e de recherche de l'Ecole Normale Sup\'erieure et de l'Universit\'e
Pierre et Marie Curie, associ\'ee au CNRS.}


\begin{references}
\bibitem{note}
We solve Eq.(\ref{eq:GPE}) numerically, using cylindrical symmetry,
on a finite grid with a splitting technique.
Numerical accuracy is
tested by changing the grid size and checking energy conservation ($\delta E/E
\leq 10^{-5}$ for $\omega t=100$).
\end{references}
\end{document}